\documentclass{emulateapj}
\usepackage[hypertex]{hyperref}
\newcommand{\adsurl}[1]{\href{#1}{ADS}}

\providecommand{\url}[1]{\href{#1}{#1}}

\def\msun{M_{\odot}}
\def\beq{\begin{equation}}
\def\eeq{\end{equation}}
\def\bea{\begin{eqnarray}}
\def\eea{\end{eqnarray}}
\def\eg{e.g.,~}

\usepackage{graphicx}
\usepackage{amssym,aas_macros}

\begin{document}
\title{Fundamental Plane of Sunyaev-Zel'dovich clusters}

\author{Niayesh Afshordi\altaffilmark{1,2}}
\email{nafshordi@perimeterinstitute.ca}
\altaffiltext{1}{Perimeter Institute
for Theoretical Physics, 31 Caroline St. N., Waterloo, ON, N2L 2Y5,
Canada}
\altaffiltext{2}{Institute for Theory and Computation, Harvard-Smithsonian Center for
Astrophysics, MS-51, 60 Garden Street, Cambridge, MA 02138, USA}


\def\LaTeX{L\kern-.36em\raise.3ex\hbox{a}\kern-.15em
    T\kern-.1667em\lower.7ex\hbox{E}\kern-.125emX}

\newtheorem{theorem}{Theorem}[section]

\begin{abstract}
Sunyaev-Zel'dovich (SZ) cluster surveys are considered among the
most promising methods for probing dark energy up to large
redshifts. However, their premise is hinged upon an accurate
mass-observable relationship, which could be affected by the (rather
poorly understood) physics of the intracluster gas. In this letter,
using a semi-analytic model of the intracluster gas that
accommodates various theoretical uncertainties, I develop a {\it
Fundamental Plane} relationship between the observed size, thermal
energy, and mass of galaxy clusters. In particular, I find that $M
\propto (Y_{SZ}/R_{SZ,2})^{3/4}$, where $M$ is the mass, $Y_{SZ}$ is
the total SZ flux or thermal energy, and $R_{SZ,2}$ is the SZ
half-light radius of the cluster. I first show that, within this
model, using the Fundamental Plane relationship reduces the
(systematic+random) errors in mass estimates to $14\%$, from $22\%$
for a simple mass-flux relationship. Since measurement of the
cluster sizes is an inevitable part of observing the SZ clusters,
the Fundamental Plane relationship can be used to reduce the error
of the cluster mass estimates by $\sim 34\%$, improving the accuracy
of the resulting cosmological constraints without any extra cost. I
then argue why our Fundamental Plane is distinctly different from
the virial relationship that one may naively expect between the
cluster parameters. Finally, I argue that while including more
details of the observed SZ profile cannot significantly improve the
accuracy of mass estimates, a better understanding of the impact of
non-gravitational heating/cooling processes on the outskirts of the
intracluster medium  (apart from external calibrations) might be the
best way to reduce these errors.

\end{abstract}

\section{Introduction}

The origin of the present-day acceleration of the Universe is
arguably the most central question in modern cosmology, and is thus
likely to dominate theoretical and observational efforts in
cosmology for decades to come.  As recently highlighted by the {\it
Dark Energy Task Force Report} \citep{2006astro.ph..9591A}, one of
the most promising methods for probing the history of cosmic
acceleration, or its most likely culprit, Dark Energy, is the
abundance of galaxy clusters at large redshifts, which is
exponentially sensitive to the cosmic expansion history
\citep{1996A&A...314...13B,1998ApJ...504....1B}. This has motivated
many upcoming cluster surveys such as APEX, ACT, SPT, and SZA, which
use the thermal Sunyaev-Zel'dovich (SZ) signature \citep{sunyaev72,
carlstrom_etal02} of the hot intracluster gas in the microwave sky
to find clusters at high
redshifts\footnote{\href{http://bolo.berkeley.edu/apexsz/}{http://bolo.berkeley.edu/apexsz/};
  \href{http://www.physics.princeton.edu/act/}{http://www.physics.princeton.edu/act/};
  \href{http://spt.uchicago.edu/}{http://spt.uchicago.edu/};
  \href{http://astro.uchicago.edu/sza/}{http://astro.uchicago.edu/sza/}}.

However, the accuracy of any cosmological constraint inferred from a
cluster survey is hinged upon how well the mass of the clusters can
be estimated from the individual cluster observables. For example,
\citet{2005JCAP...12..001F} show that a $10\%$ systematic error in
the mass estimates is enough to significantly affect the accuracy of
predicted dark energy constraints from upcoming SZ cluster surveys.
Although the total SZ flux of a cluster, which traces the total
thermal energy of the Intracluster Medium (ICM), is predicted to be
a robust tracer of its mass
\citep[\eg][]{1999PhR...310...97B,carlstrom_etal02,
2006ApJ...651..643R}, recent X-ray and SZ observations indicate that
a significant fraction of cluster baryons may have been removed from
the ICM, introducing a new uncertainty into the theoretical
predictions \citep{2006ApJ...640..691V,
2005ApJ...629....1A,2006astro.ph.12700A,2006ApJ...652..917L,
2007astro.ph..2241E}. Although self-calibration methods, through use
of phenomenological/physical ICM models \citep{2004ApJ...613...41M,
2006ApJ...653...27Y}, clustering of clusters
\citep{2004PhRvD..70d3504L,2005PhRvD..72d3006L},  or gravitational
lensing \citep{2006ApJ...649..118S,2007astro.ph..1276H} have been
put forth as a way to avoid theoretical uncertainties, they do rely
on ad hoc power-law fitting formulae and/or modeling assumptions
that could jeopardize the accuracy of their applications.

In this letter, I advocate a way to improve the accuracy of mass
estimates (or alternatively relax modeling assumptions) through
including more information about the observed SZ profile. In
particular, while the usual mass estimates only rely on the total SZ
flux, I develop a {\it Fundamental Plane} relationship
\citep{2002ApJ...581....5V} among the cluster mass, the total SZ
flux, and the SZ half-light radius of the cluster. The latter is an
independent observable for a moderately resolved cluster, and should
be readily measurable at similar precisions to the SZ flux, for the
upcoming SZ cluster surveys.

\section{Semi-Analytic model of the Intracluster Medium}

In order to study the scaling of different ICM observables, we first
develop a semi-analytic ICM model which accommodates a generous
allowance for different theoretical uncertainties. The main
ingredient in our semi-analytic ICM model is the assumption of hot
gas sitting in hydrostatic equilibrium in a nearly spherical dark
matter halo. The dark matter profile is approximated by an NFW
profile \citep{nfw}: \beq \rho(r) =
\frac{\rho_s}{(r/r_s)(1+r/r_s)^2}, \eeq where $r_s$ quantifies the
scale at which the slope of the density profile changes from $-1$ to
$-3$. This scale is often parameterized using the concentration
parameter, $c_{200}= r_{200}/r_s$, where $r_{200}$ is the radius
within which, the mean density of cluster is $200$ times the {\it
critical density} of the Universe. We assume a log-normal
distribution for $c_{200}$ with the mean: \beq \langle
c_{200}\rangle = 3.35 \left(M_{200}\over 10^{14}
h^{-1}\msun\right)^{-0.11}, \eeq and a $22\%$ scatter
\citep{2004A&A...416..853D}, which is appropriate for an
$\Omega_m=0.3$ and $\sigma_8 =0.8$ cosmology \citep[\eg see Fig. 11
in][]{2006MNRAS.372..758M}. The NFW gravitational potential can then
be derived analytically: \beq \phi(r)=
-\frac{GM_{200}}{r}\frac{\ln(1+r/r_s)}{\ln(1+c_{200})-c_{200}/(1+c_{200})}.
\eeq

Next, we populate this potential with a polytropic gas, i.e. $P_{\rm
gas} \propto \rho_{\rm gas}^{\Gamma}$, with $\Gamma \simeq 1.2$
\citep{1998ApJ...509..544S,
2005ApJ...629....1A,2005ApJ...634..964O}. Such a polytropic
distribution is expected from a turbulent rearrangement and is
roughly consistent with hydrodynamical simulations
\citep{2005ApJ...634..964O} and X-ray observations \cite[\eg][and
references therein]{2003ApJ...593..272V}. We allow a range \beq 1.1
< \Gamma < 1.3, \eeq with a flat prior, to accommodate uncertainties
in and deviations from a polytropic distribution.

In addition to thermal gas pressure, hydrostatic support can be
provided by non-thermal sources of pressure. For example,
\citet{2007ApJ...655...98N} show that subsonic turbulent pressure
can yield $5\%-20\%$ increase in pressure gradients. Moreover,
\citet{2006astro.ph.11037P} argue that cosmic rays can contribute up
to $32\%$ of the total pressure in a realistic cluster simulation.
To include this uncertainty, we consider a wide range of \beq 5\% <
 \delta_{\rm nth} < 50\%, \eeq with a flat prior,
where $\delta_{\rm nth} \equiv P_{\rm nth}/P_{\rm gas}$, is the
ratio of non-thermal to thermal pressure components \footnote{Note
that $\delta_{\rm nth}$ is not expected be constant across the ICM
\citep[\eg][]{2006astro.ph.11037P,2007ApJ...655...98N}. However, the
wide range of uncertainty that is already assumed here for
$\delta_{\rm nth}$ should also include the consequences of its
non-uniformity.}

Plugging all sources of pressure into the equation of hydrostatic
equilibrium, and using the polytropic relation, we find the ICM
temperature profile \citep[\eg][]{2005ApJ...634..964O}: \beq T(r)=
-\left(\frac{1-\Gamma^{-1}}{1+\delta_{\rm
nth}}\right)\left[\phi(r)-\phi(r_{200})\right]+T(r_{200}),\label{t_prof}\eeq
where $T(r_{200})$ is an integration constant, which is proportional
to the surface pressure of the region within $r_{200}$. We quantify
this constant through the quantity $b^2_T$
\citep{2007astro.ph..2241E}, which is defined as: \beq b^2_T =
\frac{\langle T \rangle(<r_{200})}{\mu m_p \langle \sigma^2_{\rm
DM}\rangle(<r_{200})}, \label{b2t}\eeq where $\langle T \rangle$ is
the mean gas mass weighted temperature, $\mu m_p$ is the mean
particle mass in the ICM plasma ($\mu \simeq 0.59$), and $\langle
\sigma^2_{\rm DM}\rangle$ is the mean 1D dark matter velocity
dispersion. The latter is exquisitely constrained in
\citet{2007astro.ph..2241E} through use of a host of different dark
matter simulations: \beq \langle \sigma^2_{\rm
DM}\rangle^{1/2}(<r_{200}) = (1084  ~{\rm km/s}) \left(M_{200}\over
10^{15} h^{-1} \msun\right)^{0.336}, \eeq plus $4\%$ random scatter.
The `Santa Barbara Cluster' comparison constrains the value of
$b^2_T$ to $0.87 \pm 0.04$, for a large range of different adiabatic
simulations \citep{1999ApJ...525..554F}. More recent high resolution
simulations that include cooling and feedback effects
\citep{2006ApJ...650..538N} yield consistent values \beq b^2_T =
0.90 \pm 0.11, \eeq but with larger scatter. We will adopt the
latter range with  a Gaussian distribution.

 In order to set the normalization for the gas pressure/density, we
 have to fix the total ICM baryonic budget, which we quantify
 through $f_{\rm gas} \equiv M_{\rm gas}(<r_{200})/M_{200}$. Various
 X-ray \citep{2006ApJ...640..691V, 2007astro.ph..2241E} and SZ observations \citep{2005ApJ...629....1A,2006astro.ph.12700A,2006ApJ...652..917L},
 as well as hydrodynamical simulations \citep{2006ApJ...650..538N} have
 indicated that $f_{\rm gas}$ may be significantly lower than the total
 cosmic baryonic budget, which we set to $\Omega_b/\Omega_m = 0.168$
 \citep{2006astro.ph..3449S}. To accommodate this, we also assume a
 generous range of:
 \beq 0.6 < \frac{f_{\rm gas}}{\Omega_b/\Omega_m} < 0.9, \eeq
for ICM gas mass fractions.

To estimate the total ICM SZ flux, we need to know the outer edge of
our ICM model, or the radius of the accretion shock, $r_{\rm max}$.
Assuming that gas comes to stop at the shock, the temperature behind
the shock is roughly given by \beq T(r_{\rm max}) \simeq \frac{1}{3}
\mu m^2_p v^2_{\rm inf.} (1+\delta_{\rm nth})^{-1},
\label{tshock}\eeq where $v_{\rm inf.}$ is the gas infall
velocity\footnote{Here we have assumed that the non-thermal pressure
component behaves as non-relativistic monatomic gas, which is not
appropriate for cosmic ray pressure. However, we will ignore this
difference in our model.}. We then use the value of infall velocity
from the spherical collapse model \citep{1972ApJ...176....1G}: \beq
v_{\rm inf.} = \sqrt{GM(r)\over r} \left\{\sqrt{2}
-[2\Delta_c]^{-1/3} (Ht/\pi)^{-2/3} + O(\Delta_c^{-2/3})\right\},
\label{infall}\eeq where $\Delta_c$ is the mean overdensity with
respect to the critical density within the shock radius. Combining
Eqs. (\ref{t_prof}-\ref{b2t}, \ref{tshock}-\ref{infall}) with mean
densities from the NFW profile fixes the outer edge of our ICM
model.

The final step is to include the ellipticity/triaxiality of real
haloes in our model. The impact of triaxiality on the total SZ flux
of a halo is of second order, and so we will neglect it in our
analysis. However, triaxiality introduces a random scatter in the
projected SZ profiles, which will impact the observed half light
radii. To model this, we assume that, to first order, the triaxial
profile has the shape: \beq P(r,\theta,\varphi) =
\bar{P}(r)\left[1+\sum_{m=-2}^2 a_{2m}
Y_{2m}(\theta,\varphi)\right], \eeq where $\bar{P}(r)$ is the
prediction from our spherical polytropic model, and $Y_{2m}$'s are
spherical harmonics. We then assume a Gaussian distribution with a
reasonable range of \beq \langle a^2_{2m}\rangle^{1/2} = 0.16,
\langle a_{2m} \rangle =0, \eeq
 to model the triaxiality of real clusters. This amplitude of
 triaxiality has equivalent moments to an ellipsoidal distribution
 with axes ratios of 1:0.7:0.5 \citep[expected for CDM haloes,
 \eg][]{1991ApJ...378..496D}, and a (spherically averaged) pressure
 profile $P(r) \propto r^{-2}$, which is roughly consistent with
 observations and simulations of SZ clusters \citep{2006astro.ph.12700A}.

\section{Fundamental Plane of SZ clusters}

Let us first quantify the SZ flux of a cluster in terms of $Y_{SZ}$,
which we define as \beq Y_{SZ} \equiv {\rm Flux(mK\cdot arcmin^2)}
\left[H(z) d_A(z)/c\right]^2 h^{-1}(z), \eeq where ${\rm
Flux(mK\cdot arcmin^2)}$ is the total observed cluster SZ flux at
low frequencies, in units of ${\rm mK\cdot arcmin^2}$, while $H(z) =
100 h(z)~ {\rm km/s/Mpc}$ and $d_A(z)$ are the Hubble constant and
the angular diameter distance at redshift $z$. $Y_{SZ}$ can then be
easily described in terms of the properties of the ICM model: \beq
Y_{SZ} = 1.022 \left(\frac{M_{\rm gas} h(z)}{10^{15}
\msun}\right)\langle T({\rm keV}) \rangle. \eeq

Now we can generate a random set of 3000 clusters uniformly
distributed in the range \beq 13 < \log(M_{200} h/\msun) < 16, \eeq
with their ICM properties according to the prescription that we
outlined above \footnote{Notice that none of the assumption that
have gone into our ICM model would cause a break in the slope(s) of
the resulting scaling relations, and so the range assumed for
cluster masses does not change the slope or scatter of the scaling
relations.}. This leads to our mass-SZ flux scaling relation:

\beq M_{200} = (8.1 \times 10^{14} \msun/h)~ Y^{0.58}_{SZ}
~~~~{\rm(\pm 22\% ~error)}.\label{m-sz}\eeq

The error quoted here is the r.m.s. scatter around our best fit
scaling relation, and reflects a very conservative estimate of all
the theoretical uncertainties in mass-SZ flux relation. Also notice
that this includes both systematic and random uncertainties, which
cannot be distinguished in our approach. A further simplification is
the assumed lack of covariance between different uncertainties.
While possible constructive/destructive covariances could lead to
larger/smaller scatter, their correct account would require a more
detailed understanding of the various involved processes.

\begin{figure}
\includegraphics[width=\linewidth]{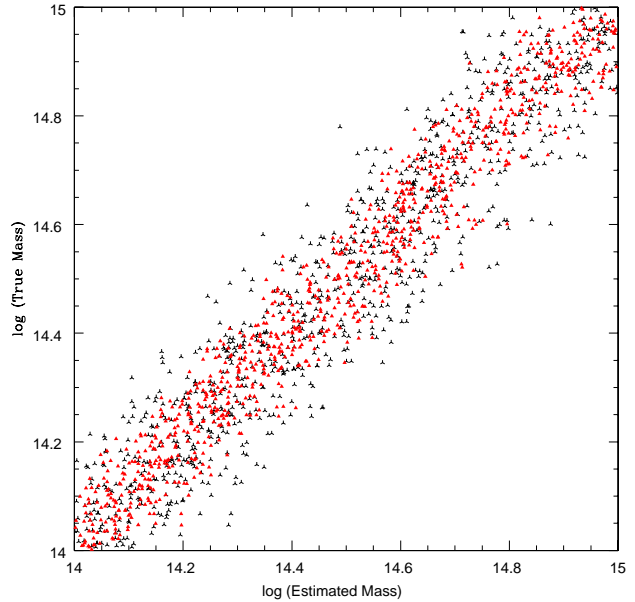}
\caption{\label{szplane} Contrast between the usual SZ flux mass
estimates, Eq. (\ref{m-sz}), shown by starred (black) points and
Fundamental Plane mass estimates, Eq. (\ref{fp}), shown by solid
(red) triangle. The error decreases from 22\% for the former, to
14\% for latter mass estimates.}
\end{figure}

To approach the main subject of this letter, i.e. the Fundamental
Plane of SZ clusters, we will next include information about the ICM
SZ profile into our scaling relation. We do this by calculating the
{\it half-light radius}, $R_{SZ,2}$, which is defined as the radius
of the disk (or cylinder) that contains half of the total SZ flux.
The new best fit scaling relation for our clusters is: \bea M_{200}
= (7.8 \times 10^{14} \msun/h)~Y^{0.75}_{SZ} \left(R_{SZ,2}\over{\rm
Mpc}/h\right)^{-0.76}\nonumber\\ {\rm(\pm 14\%
~error),}\label{fp}\eea which shows an almost $34\%$ decrease in the
error of the mass estimate. The new scaling relation is shown by
solid (red) triangles in Figure (\ref{szplane}), which should be
contrasted with the starred (black) points that result from the
usual mass-SZ flux relation (Eq. \ref{m-sz}).

One may wonder if using more information about the SZ profile of the
cluster can help reduce the errors in mass estimates even further.
To investigate this, we can add the radius of the disk containing a
quarter of the total projected SZ flux, $R_{SZ,4}$, into the list of
observables. However, we find that the resulting estimator, which
now depends on $Y_{SZ}$, $R_{SZ,2}$, and $R_{SZ,4}$, has a scatter
of $13\%$, which is almost the same as the error in the Fundamental
Plane estimator. Therefore, we conclude that adding more details
about the SZ profile is unlikely to improve the accuracy of mass
estimates significantly.

\section {Physical Origin of the Fundamental Plane}

It is interesting to notice that our Fundamental Plane relationship
(Eq. \ref{fp}) is different from the virial relation, $M \propto
(Y_{SZ}R_{SZ,2})^{1/2}$, previously adopted \eg in
\citet{2002ApJ...581....5V}. The reason for this apparent
discrepancy is that the virial relation is only an approximation
which results from the assumption of hydrostatic equilibrium and
self-similar pressure profiles for different clusters. More specific
information about the initial conditions of the cosmological
collapse, or the surface pressure, while relaxing the
self-similarity assumption, can lead to more accurate scaling
relations. Since we use the assumption of hydrostatic equilibrium,
most of our clusters sit close to the intersection of the virial
relation and Eq. (\ref{fp}), but (by construction) are better fit by
Eq. (\ref{fp}).

There is a simple way to understand the physical origin of Eq.
(\ref{fp}) analytically. In Sec. \ref{sec_errors}, we show that
(within our model) the scatter in the mass-flux relation  is
dominated by the uncertainty in the surface pressure, which sets the
outer boundary of the ICM (Table \ref{error_sources}). If we
approximate the gravitational potential within the ICM region that
dominates the SZ flux by an isothermal potential, and fix all the
cluster/ICM parameters, other than its outer boundary, we have
$M_{\rm gas} \propto R_{SZ,2}$, while $\langle T_{\rm gas}\rangle
\simeq ~{\rm const}$. Therefore, $Y_{SZ} \propto R_{SZ,2}$ for a
fixed cluster potential (and thus fixed $M_{200}$): \beq F(M_{200})
= Y_{SZ}/R_{SZ,2}. \eeq Combining this with the standard scaling
relations: $Y_{SZ} \propto M^{5/3}_{200}$ and $R_{SZ,2} \propto
M_{200}^{1/3}$ yields: \beq M_{200} \propto (Y_{SZ}/R_{SZ,2})^{3/4}.
\eeq

\section{Discussions}

\subsection{Breakdown of Error Budgets}\label{sec_errors}
\begin{table}
\begin{center}
\caption{\label{error_sources} Breakdown of the fractional error in
mass estimates into contributions due to different sources of
uncertainty.}

\begin{tabular}{|c|c|c|}
  \hline
    sources of $\Delta M^2/ M^2$ & in Mass-Flux & in Fundamental Plane \\
  \hline
   halo concentration & 0.001 & 0.003 \\
   polytropic index & 0.000 & 0.000 \\
   gas fraction & 0.004 & 0.008 \\
   non-thermal pressure & 0.005 & 0.004 \\
   DM velocity dispersion & 0.010 & 0.005 \\
   surface pressure & 0.025 & 0.009 \\
   triaxiality & 0.000  & 0.000 \\
   total & 0.048 & 0.022 \\
  \hline
\end{tabular}
\end{center}
\end{table}

In Table \ref{error_sources}, we have broken down the errors in mass
estimates into contributions from different sources of uncertainty.
To do this for each parameter, we assume that it only takes the
central value of its assumed distribution, and then measure the
change in the error quadratures of the new scaling relations.

 We see that uncertainties in the ICM gas fraction,
non-thermal and surface pressures, as well as the dark matter
velocity dispersion are the main sources of error in our mass
estimates. In contrast, the uncertainties in the halo concentration,
the polytropic index, the ICM outer edge, or the halo triaxiality
have little impact on the errors.

In particular, the error in both scaling relations seem to be
dominated by our assumed uncertainty in the ICM surface pressure,
which is ultimately related with the amount of non-gravitational
heating/cooling associated with galaxy or black hole formation in
the clusters.

Another interesting observation is that the uncertainty in gas
fraction has a larger contribution to the error in the Fundamental
Plane relation, than to the error in the mass-flux relation. This is
due to the fact that $R_{SZ,2}$ does not contain any information
about $f_{\rm gas}$ (at least in our model), while Fundamental Plane
masses have a steeper dependence on $Y_{SZ}$, and thus are more
sensitive to $f_{\rm gas}$ uncertainty.

Notice that possible correlations among different uncertainties,
that are overlooked in our simple ICM model, may tilt the
Fundamental Plane from our Eq. (\ref{szplane}), and also change the
size of the errors. However, presence of any such correlations is
not immediately obvious in our current theoretical understanding of
the ICM physics.

One may wonder why the error in our mass-flux relationship (Eq.
\ref{m-sz}) is so much larger than those advocated in numerical
studies such as \citet{2006ApJ...650..128K}, which are only $\sim
6\%$. The reason is that these studies measure the SZ flux within 3D
spheres of fixed overdensity ($= 500$ times critical, for
\citet{2006ApJ...650..128K}), while our total SZ flux is integrated
out the ICM accretion shock. Of course, the latter is a more
relevant quantity for 2D SZ cluster observations, especially for
poor angular resolutions. In fact, our $M_{500}-Y_{SZ,500}$ relation
has only a scatter of $9\%$, which is reasonable as we include more
theoretical uncertainties than in \citet{2006ApJ...650..128K}'s
simulations. This shows that the bulk of the scatter in our
mass-flux relationship comes from the uncertainty in the outer edge
of our ICM model.

Finally, we should point out that a breakdown into systematic and
random errors is also not possible within our exercise, due to our
poor statistical understanding of different non-gravitational
processes (such as cosmic ray injection or stellar feedback) that
affect the scaling relations.
\subsection{Noisy Observations}
As the purpose of this letter is to introduce a novel and improved
mass estimator for SZ cluster surveys, we defer a detailed study of
the observational issues associated with the use of this method to
future investigation. Such details, while important, should be
suited to the specifics of each survey, as well as the class of
cosmological models that one would intend to constrain. However, in
what follows, I will outline some of the steps that need to be taken
for a realistic cosmological application.

We should first recognize that any realistic observation of SZ
clusters is limited both by the finite detector noise, as well as
the finite beam resolution. While a poor resolution does affect the
precision of the SZ flux measurement, its impact is much more severe
for the cluster size measurement. For example, if the detector beam
is significantly larger than the virial radius of the cluster, then
the Fundamental Plane relation cannot add much to the mass-flux
relation, even if the cluster is detected at several-$\sigma$ level.

In the absence of perfect resolution, the most practical way to use
the Fundamental Plane relation is to fit a parametrized template
(\eg a Gaussian) to the observed cluster SZ map, and replace
$R_{SZ,2}$ with the characteristic scale of the template,
$\sigma_{SZ}$ \footnote{Of course, the normalization/slopes of the
scaling relations should be re-calculated for the specific template.
Here, we assume that the Fundamental Plane or its scatter would not
change significantly.}. Assuming both Gaussian template and beam,
this measurement is done by minimizing the following $\chi^2$
function: \beq \chi^2 = \int\frac{d^2\ell}{(2\pi)^2}
\frac{\left[T_{obs,\vec{\ell}}-Y_{SZ}
e^{-\ell^2(\sigma^2_{SZ}+\sigma^2_{\rm
beam})/2}\right]^2}{C_{\ell}e^{-\ell^2\sigma^2_{\rm beam}}+N},
\label{chi2}\eeq where $T_{obs,\vec{\ell}}$ is the flat-sky Fourier
transform of the cluster SZ map, $C_{\ell}$ is the CMB power
spectrum, $N$ characterizes the detector noise, and $\sigma_{\rm
beam}$ is the size of the detector beam.

In the limit that the detector noise is the primary source of
measurement uncertainty ($C_{\ell}e^{-\sigma^2_{\rm beam}\ell^2} \ll
N$), the Fisher matrix resulting from Eq. (\ref{chi2}) reduces to
Gaussian integrals which can be calculated analytically. In
particular, for a well-resolved cluster ($\sigma^2_{SZ} \gtrsim
\sigma^2_{\rm beam}$), the total (measurement+theory) error in
Fundamental Plane masses is only $\sim 28\% (22\%)$ smaller than the
SZ-flux mass estimates for a cluster that is detected at a $5\sigma
(3\sigma)$ level. This is due to the fact that the theoretical mass
degeneracy direction in the $Y_{SZ}-R_{SZ,2}$ (or
$Y_{SZ}-\sigma_{SZ}$) plane does not coincide with the degeneracy
direction of the measured parameters. Including a finite resolution
will only further deteriorate the performance of the Fundamental
Plane mass estimates.

\section{Conclusions}

To summarize, using a semi-analytic model of the intracluster medium
which accommodates different theoretical uncertainties, we found a
Fundamental Plane relationship that relates the mass of galaxy
clusters to their SZ flux and SZ half-light radius. Use of this
relationship should lead to $\sim 34\%$ smaller error in mass
estimates in comparison to the usual mass-flux relation, and hence
more accurate cosmological constraints. While including more details
about the SZ profile is unlikely to increase the accuracy of mass
estimates, a better understanding of the role of non-gravitational
heating/cooling processes should significantly reduce the errors.

While the goal of this letter was to introduce the idea of using
Fundamental Plane relationship to improve SZ cluster mass estimates,
there is much more that remains to be done in order to exploit the
full potential of this method. For example, rather than focusing on
the cluster masses, we can study the cosmological constraints from
the full bivariate distribution of cluster fluxes and half-light
radii $n(Y_{SZ},R_{SZ,2})$. This combines the traditional
mass-function constraints, with the constraints resulting from
observed scaling relations, as advocated by
\citet{2002ApJ...581....5V}. Another topic that we did not address
here was the impact of including merging (non-relaxed) clusters
and/or false detections in our sample. One may identify (and thus
exclude) these clusters as outliers in the $Y_{SZ}-R_{SZ,2}$ plane
for a given redshift bin, which would not have been possible in the
absence of SZ profile information. Finally, it is needless to say
that the true Fundamental Plane, as well as the full impact of
different theoretical uncertainties, can only be accurately (and
adequately) modeled through high-resolution and realistic
cosmological simulations of a fair sample of galaxy clusters.

\section*{Acknowledgments}

I would like to thank Daisuke Nagai, Licia Verde, Zoltan Haiman,
David Spergel, and Beth Reid for helpful comments on this
manuscript. I also would like to thank Daisuke Nagai for providing
the pressure profiles of the simulated clusters in
\citet{2006ApJ...650..538N}.

\bibliography{szplane_3}

\end{document}